\RequirePackage[2020-02-02]{latexrelease}
\documentclass[aps]{revtex4-2}
\usepackage{amssymb}
\usepackage{amsmath,amssymb,graphicx}
\usepackage{graphicx}
\usepackage{dcolumn}
\usepackage{bm}
\def\be{\begin{equation}}
\def\ee{\end{equation}}
\def\bea{\begin{eqnarray}}
\def\eea{\end{eqnarray}}

\def\lsim{\raise0.3ex\hbox{$\;<$\kern-0.75em\raise-1.1ex\hbox{$\sim\;$}}}
\def\gsim{\raise0.3ex\hbox{$\;>$\kern-0.75em\raise-1.1ex\hbox{$\sim\;$}}}

\newcommand{\mpi}{M_{\pi}}

\newcommand{\mk}{M_K}
\newcommand{\mtau}{m_\tau}

\newcommand{\pk}{p_K}
\newcommand{\ppi}{p_\pi}

\newcommand{\Lagr}{\mathcal{L}}

\usepackage{pstricks}
\begin{document}

\title{CP Asymmetry in $\tau \to K\pi \nu_{\tau} $ within Non-Minimal $SU(5)$ }

\author{Gaber Faisel}
\email{gaberfaisel@sdu.edu.tr}

\affiliation{{\fontsize{10}{10}\selectfont{Department of Physics,
Faculty of Arts and Sciences, S\"uleyman Demirel University,
Isparta, Turkey 32260.}}}

\author{ S. Khalil}
\email{skhalil@zewailcity.edu.eg}

\affiliation{{\fontsize{10}{10}\selectfont{Center for Fundamental Physics, \\
Zewail City of Science and
Technology, 6th of October City, Giza 12578, Egypt.}}}

\begin{center}
\begin{abstract}

Direct CP asymmetry in semi-leptonic $\tau$ decays is an
intriguing hint for new physics beyond the standard model. We
investigate the CP asymmetry in $\tau^- \to K_S^0 \pi^- \nu_\tau$
and $\tau^- \to K^- \pi^0 \nu_\tau$ decays in non-minimal SU(5)
model, with 45-dimensional Higgs multiplet. We show that the
associate color-triplet scalar is a natural example of a scalar
leptoquark that mediates the transition $\tau \to s \bar u
\nu_\tau$, and accounts for the 2.8 $\sigma$ discrepancy between
the $A_{\rm CP}(\tau^- \to K_S^0 \pi^- \nu_\tau)$ experimental
results and the SM expectation. Furthermore, it predicts a sizable
$A_{\rm CP}(\tau^- \to K^- \pi^0 \nu_\tau)$, of order ${\cal
O}(10^{-3})$, which can be accessible in current and near future
experiments.

\end{abstract}
\end{center}

\maketitle

\section{Introduction}
Semi-leptonic $\tau$ decays are very interesting venue for
investigating Physics Beyond the Standard Model (BSM). The CP
asymmetry of $\tau \to K^0_S \pi \nu_\tau$ is an example of an
observation that points to new physics. The Belle collaboration
reported evidence for CP violation in the $\tau \to K_S \pi \nu$
decay mode in 2011 \cite{Belle:2011sna}, which the BaBar
collaborations later confirmed \cite{BaBar:2011pij}. It was the
first time to observe  CP violation in a purely leptonic decay
process. The observation of direct CP violation in the $\tau \to
K^0_S \pi \nu$ decay  has sparked further research into physics
BSM. According to the result reported by BaBar collaborations, the
CP asymmetry in $\tau^- \to K_S^0 \pi^- \nu_\tau$ decay is given
by \cite{BaBar:2011pij}. \bea A_{\rm CP}(\tau^+ \to K_S^0 \pi^+
\nu_\tau) &=& \frac{\Gamma(\tau^+ \to K_S^0 \pi^+ \nu_\tau)
 - \Gamma(\tau^- \to K_S^0 \pi^- \nu_\tau)}{\Gamma(\tau^- \to K_S^0 \pi^- \nu_\tau) -
 \Gamma(\tau^+ \to K_S^0 \pi^+ \nu_\tau)}\nonumber\\
&=& - (3.6 \pm 2.3\pm 1.1) \times 10^{-3}. \eea According to the
Standard Model (SM) this process occurs via $\tau^- \to s
\bar{u}\nu_{\tau}$ transition and no direct CP violation signal is
expected. However due to the CP violation in $K^0-\bar{K}^0$
mixing at one loop level, the SM expectation of $A_{\rm CP}(\tau^+
\to K_S^0 \pi^+ \nu_\tau)$ is of order $2 \times
Re(\varepsilon_K)$, and the total signal should be
\cite{Grossman:2011zk,Bigi:2005ts,Calderon:2007rg,Chen:2019vbr}
\be A_{\rm CP}(\tau^+ \to K_S^0 \pi^+ \nu_\tau)\vert_{\rm SM} =
(3.32 \pm 0.06) \times 10^{-3}. \ee Therefore, there is a 2.8
sigma discrepancy that may indicate the presence of direct CP
violation, which is absent in the SM. The presence of new sources
of CP violation beyond the Cabibbo-Kobayashi-Maskawa (CKM) matrix
in BSM can contribute to the $\tau \to K^0_S \pi \nu$ decay's
direct CP asymmetry. These contributions could result from new
interactions involving particles not found in the standard model,
such as charged Higgs boson, new particles in supersymmetric
extensions of the SM or the scalar leptoquark.

The decay $\tau^-\to K^- \pi^0 \nu_\tau$ has also been
investigated experimentally, with a branching ratio of order $2.7
\times 10^{-3}$ \cite{Workman:2022ynf}. There have been attempts
to measure the direct CP asymmetry of this decay, motivated by
observations of CP violation in the $\tau^- \to K_S^0 \pi^-
\nu_\tau$ decay channel. However, no reported results have been
provided so far. Because these two processes are correlated, we
will predict the direct CP asymmetry in the $K^-$ channel
associated with accounting for the $2.8$ discrepancy in $K_S^0$
channel, by the new physics CP violation effect.

Non-vanishing direct CP asymmetry requires an interference between
two contributed amplitudes with non-vanishing weak and strong
relative phases. If the total amplitude is given by $A_1 e^{i
\delta_1^s} e^{i \delta_1^w} + A_2 e^{i \delta_2^s} e^{i
\delta_2^w}$, then the CP asymmetry is given by \be A_{\rm CP} = -
4 \vert A_1\vert \vert A_2 \vert \sin \delta^s \sin\delta^w, \ee
where $\delta^s =\delta_1^s- \delta_2^s$ and $\delta^w= \delta_1^w
- \delta_2^w$. Both $K^-$ and $K_S^0$ channels occur on quark
level in the SM via the tree level transition $\tau \to s \bar u
\nu_\tau$, so their amplitudes are essentially real. After going
beyond the tree level, the SM predicts negligibly direct CP
asymmetry of order $10^{-12}$ \cite{Delepine:2005tw}. Previous
studies of CP violation in this decay channel has been conducted
in supersymmetric extension of the SM
\cite{Delepine:2006fv,Delepine:2007qg}, multi Higgs models with
complex couplings \cite{Delepine:2018amd} and effective models
with scalar leptoquarks \cite{Delepine:2018amd}. These studies
demonstrated that the estimated direct CP asymmetry is so small
that further research is warranted.

The scalar leptoquark is one of the best candidates for mediating
this decay at the tree level. It contributes to the amplitude with
scalar and tensor operators that have non-vanishing strong and CP
weak phases that differ from the SM CKM phase. Furthermore,
leptoquark is naturally predicted in GUT extensions of the SM. One
version of the $SU(5)$ GUT model includes $45$ (adjoint) Higgs
fields \cite{Georgi:1974sy}. This model is known as the
"non-minimal $SO(5)$. The adjoint Higgs field contributes to
$SU(5)$ breaking down to the SM gauge group and provides a good
fit to the observed particle masses and mixing angles, thereby
resolving one major problem with the minimal $SU(5)$. The
color-triplet scalar is one of the components of the $45$ Higgs,
which can be as light as TeV if fine tuning similar to the
well-known doublet-triplet splitting in GUT models is considered
\cite{Khalil:2013ixa}. It is worth noting that because it has no
coupling with quark-quark \cite{LPTOdec}, this triplet does not
contribute to proton decay.

In this paper, we consider the above-mentioned $SU(5)$ color-
triplet scalar, which is a natural example of scalar leptoquark.
This type of leptoquark differs from the general adhoc examples of
leptoquarks discussed  in the literature. As previously stated, it
interacts in a specific way, so it does not contribute to proton
decay but does contribute to $\tau \to s \bar u \nu$ decay.   We
emphasize that this contribution can account for the observed
direct CP asymmetry in $\tau \to K \pi \nu$ decay.

The paper is structured as follows. Section 2 provides a brief
overview of the CP Asymmetry in $\tau \to K \pi \nu$ decay within
the SM and beyond. Section 3 is devoted to discussing the $SU(5)$
scalar leptoquark and its associated interactions, emphasizing
that while it does not contribute to proton decay, it can play a
significant role in the decays under consideration. The analysis
of the new contribution of our scalar leptoquark to direct CP
asymmetries of the decay modes $\tau^- \to K^0_S \pi^- \nu$ and
$\tau^- \to K^- \pi^0 \nu$  is discussed in section 4. Finally our
conclusions and prospects are give in section 5.

\section{CP Asymmetry in $\tau \to K \pi \nu$ decay
\label{Asy}}

The effective Hamiltonian of $\vert \Delta S \vert=1$ $\tau $
decays is given by \be {\mathcal H}_{eff} =
-\frac{G_F}{\sqrt{2}}V^\star_{us}\sum_{i=V, A,S,P,T}
C_i(\mu)\,Q_i(\mu),\label{Hef}\ee where $V_{us}$ is the CKM mixing
matrix element and $Q_i$ represent the four-fermion local
operators at low energy scale $\mu\simeq m_\tau$ where \bea Q_{V}
&=&\big(\bar{\nu}_\tau \gamma_\mu\tau\big)\big(\bar{s} \gamma^\mu
u\big),
~~~~~~~~~  Q_{A} =\big(\bar{\nu}_\tau \gamma_\mu \gamma_5 \tau\big)\big(\bar{s} \gamma^\mu  u\big),\nonumber\\
Q_{S} &=& \big(\bar{\nu}_\tau  \tau\big)\big(\bar{s}  u\big),
~~~~~~~~~~~~~~~~  Q_{P} = \big(\bar{\nu}_\tau \gamma_5 \tau\big)\big(\bar{s}  u\big),\nonumber\\
Q_{T} &=& \big(\bar{\nu}_\tau \sigma_{\mu \nu } (1+ \gamma_5) \tau
\big)\big(\bar{s}\sigma^{\mu \nu } u\big),\label{Qi}\eea with
$\sigma_{\mu\nu}= \frac{i}{2} [\gamma_\mu, \gamma_\nu]$. The
Wilson coefficients, $C_i$, corresponding to the operators $Q_i$
can be expressed as \be C_i = C^{SM}_i+C^{NP}_i, \ee  where
$C^{SM}_i$ and $C^{NP}_i$ represent SM and NP contributions to the
Wilson coefficients respectively. The Wilson coefficients $C_i$
are typically expressed in terms of three independent
coefficients: $C_V$ (dominated by SM contribution), $C_S$, and
$C_T$. The matrix elements of the vector, scalar and tensor
quark currents in the operators listed in Eq.(\ref{Qi}) relevant
to the process $\tau^- \to K^- \pi^0 \nu_\tau$, can be expressed
as:
\bea
 \langle K^- \pi^0|\bar s\gamma^\mu u|0\rangle&=& \frac{1}{\sqrt{2}}
 \bigg((\pk-\ppi)^\mu f_+(s)+(\pk+\ppi)^\mu f_-(s)\bigg),\nonumber\\
 \langle K^- \pi^0|\bar s u|0\rangle &=&
\frac{(\mk^2-\mpi^2)} {\sqrt{2}(m_s-m_u)}f_0(s) =
\frac{\Delta^2_{K\pi}}{\sqrt{2}(m_s-m_u)}f_0(s), \nonumber\\
\langle K^- \pi^0|\bar s\sigma^{\mu\nu} u|0\rangle &=&
\frac{i(\pk^\mu\ppi^\nu-\pk^\nu\ppi^\mu)}{\sqrt{2}\mk}B_T(s).
\label{f3} \eea The matrix elements of the decay $\tau^- \to K^0_S
\pi^- \nu_\tau$, on the other hand, can be obtained by multiplying
the right hand side of the corresponding ones in Eq.(\ref{f3}) by
$\sqrt{2}$. As shown in Ref.\cite{Delepine:2018amd}, the
differential decay width of $\tau^- \to K^- \pi^0 \nu_\tau$ is
given by
\begin{eqnarray}
\frac{d\Gamma}{d s}&=& G_F^2|V_{us}|^2 |C_V|^2 S_{\rm EW}
\frac{\lambda^{1/2}(s,\mpi^2,\mk^2)(\mtau^2-s)^2 \Delta^4_{K\pi} }{1024\pi^3\mtau s^3}\nonumber\\
&\times&\bigg[\frac{(\mtau^2+2s)\lambda(s,\mpi^2,\mk^2)}{3\mtau^2
\Delta^4_{K\pi} }\bigg(|f_+(s)
-T(s)|^2+\frac{2(\mtau^2-s)^2}{9s\mtau^2}|T(s)|^2\bigg)+|S(s)|^2\bigg],
\label{decay_width}
\end{eqnarray}
where $\lambda(x,y,z)$ is given by $\lambda(x,y,z) = x^2+y^2 +z^2
-2 xy -2 xz -2 yz$. Here $s$ is the invariant mass of the $\pi K$
system defined as $s=(\pk+\ppi)^2$,  $S_{\rm EW}=1.0194$ accounts
for the electroweak running down to $\mtau$, and  $\Delta^2_{K\pi}
= \mk^2-\mpi^2$.  The quantities $S(s)$ and $T(s)$ are defined as
\begin{eqnarray}
S(s)&=& f_0(s)\bigg(1+\frac{s\, C_S}{\mtau(m_s-m_u)\, C_V} \bigg),\nonumber\\
T(s)&=&\frac{3s}{\mtau^2+2s}\frac{\mtau \, C_{T}}{\mk C_V} \,
B_T(s).
\label{VT1}
\end{eqnarray}
It should be noted that the
differential decay width of $\tau^- \to K^0_S \pi^- \nu_\tau$ is
twice that one of $\tau^- \to K^- \pi^0 \nu_\tau$ due to the
difference in their form factors by a factor $\sqrt{2}$. The decay
rates $\Gamma(\tau^- \to K_S^0 \pi^- \nu_\tau)$ and $\Gamma(\tau^-
\to K^- \pi^0 \nu_\tau)$  can be obtained after integrating the
differential decay widths  with respect to the kinematic variable
$s$. This allows us to have a prediction of the direct CP
asymmetry of the given decay mode. In fact, as  concluded in
Refs.\cite{Cirigliano:2017tqn,Delepine:2018amd}, this asymmetry
can be generated through the interference between the SM vector
operator and new physics tensor operator, while the new physics
scalar operator does not contribute to the asymmetry. After fixing
the matrix elements and other involved parameters with their
central values, one finds that
\cite{Cirigliano:2017tqn,Delepine:2018amd}\bea
|A_{CP}(\tau^- \to K_S^0 \pi^- \nu_\tau)| & \lsim& 0.03 \times | {\rm Im}(C_T)|,\nonumber\\
|A_{CP}(\tau^- \to K^- \pi^0 \nu_\tau)| & \lsim& 0.014 \times  |
{\rm Im}(C_T)|, \label{asym}
\eea


\section{$SU(5)$ Scalar Leptoquark and its Role in $\tau \to K_S \pi \nu$}

As previously advocated, extending the Higgs sector of $SU(5)$  by
$45_H$ helps to solve some of the problems that this simple
example of GUT model faces \cite{GUT, gut, khalil, KHALIL}.
The $45_H$ transforms under the SM gauge as%
\bea %
45_H = (8,2)_{1/2}\oplus (1,2)_{1/2}\oplus (3,1)_{-1/3} \oplus
(3,3)_{-1/3}
 \oplus  (6^*,1){-1/3}\oplus (3^*,2)_{-7/6}\oplus
(3^*,1)_{4/3}.%
\eea%
It also satisfies the following constraints: $45^{\alpha
\beta}_\gamma = - 45^{\beta \alpha}_\gamma$ and $\sum_\alpha^5
(45)^{\alpha \beta}_\alpha =0$. Through non-vanishing Vacuum
Expectation Values (VEVs) of $5_H$ and $45_H$: $\langle 5_H
\rangle = v_5,\langle 45_H \rangle^{15}_1 = \langle 45_H
\rangle^{25}_2 = \langle 45_H \rangle^{35}_3 = v_{45}, \langle
45_H \rangle^{45}_4 = -3 v_{45}$, the electroweak symmetry
$SU(2)_L \times U(1)_Y$ is spontaneously broken into $U(1)_{em}$.

The $45_H$ scalar triplets are defined as: \bea &&
(3^*,2)^{ij}_{c\ -7/6} \equiv (45_H)^{ij}_c \equiv \Phi^{ij}_c ,
\\\nonumber &&  (3^*,1)^{a b}_{k\ 4/3} \equiv (45_H)^{ab}_k \equiv  \Phi^{ab}_k ,
\\\nonumber && [(3,1)^{ib}_c \oplus (3,3)^{ib}_c]_{-1/3} \equiv (45_H)^{ib}_c \equiv \Phi^{ib}_c .
\eea %
It has been emphasized \cite{LPTOdec} that while the scalar
triplets $\Phi^{ab}_k$ and $ \Phi^{ib}_c $ contribute to the
proton decay and they must be superheavy,  the scalar triplet
$\Phi^{ij}_c$ does not. It has no interaction terms that would
cause proton decay.  By writing $\Phi^{ij}_c$ as $(\phi^i_1,
\phi^i_2)^T$, one can demonstrate that the scalar triplet has the
following peculiar interactions in the weak eignestate
basis\cite{LPTOdec}: \bea {\cal L}= 2 Y'^2_{AB} e'^{T}_A C u'^c_{B
i} \phi^{i1*} + 4 (Y'^4_{AB}- Y'^4_{BA} ) u'^{iT}_A C e'^{c}_B
\phi_{i1} -2 Y'^2_{AB} \nu'^{T}_A C u'^c_{B i} \phi^{i2*}+  4
(Y'^4_{AB}-Y'^4_{BA} ) d'^{iT}_A C e'^{c}_B \phi_{i2}.~
\label{lag} \nonumber\\\eea

Working in the basis in which the weak eignestate basis $(d',u',
\nu',e' )$ are related to the mass eignestate basis $(d,u, \nu,e
)$ via the trnasformations $d'_A \to V^{CKM}_{AB} d_B,~~ \nu'_A
\to V^{\rm PMNS}_{AB} \nu_B,~~ u'_A \to u_A,~~ e'_A \to e_A,$ and
defining $Y^2_{AB}=Y'^{2}_{AB}$, $Y^{4}_{AB}
\equiv(Y'^4_{AB}-Y'^4_{BA})$ results in the Lagrangians expressing
the Yukawa couplings describing the SM fermions interactions with
the scalar leptoquark $\phi_{i2}$ and $\phi_{i1}$
\bea \mathcal{L}_{ \phi_{i1}} &=& -2Y^2_{ AB}\bar{u}_{Bi}P_{L}
e_{A}\phi^{i1*} - 4Y^{4}_{ AB}\bar{e}_{B}P_{L}
u_{A}\phi_{i1}\!-2Y^{2\,*}_{ AB}\bar{e}_{A}P_{R} u_{Bi}\phi^{i1} -
4Y^{4\,*}_{
AB}\bar{u}_{A}P_{R} e_{B}\phi^*_{i1}\!,\nonumber\\
\mathcal{L}_{ \phi_{i2}} &=& 2Y^2_{ AB} V_{ AK}^{\rm
PMNS}\bar{u}_{Bi}P_{L} \nu_{k}\phi^{i2*} - 4Y^{4}_{ D B} V^{\rm
CKM}_{ D K} \bar{e}_{B}P_{L} d_{K}\phi_{i2}\!+2Y^{2\,*}_{
AB} V_{ AK}^{\rm PMNS*} \bar{\nu}_K P_{R} u_{Bi}\phi^{i2}\nonumber\\
&-& 4Y^{4*}_{ D B}V^{\rm CKM*}_{ D K} \bar{d}_{K}P_{R}
e_{B}\phi^*_{i2}\!,\label{lag1} \eea where $P_{R,L}= (1\pm
\gamma^5)/2$ and we have used $C^T=-C$ and
$\bar{\Psi}=\Psi_{L}^{cT}$. Clearly from the previous equation
that the scalar leptoquarks $\phi_{i1}$ and $\phi_{i2}$ have
electric charges $-5/3$ and $-2/3$ respectively. As can be seen
also from the Lagrangians in Eq.(\ref{lag1}) that, $\tau^-\to
s\bar u\nu_\tau$ transition relevant to our process can be
generated by integrating out the scalar leptoquark $\phi_{i2}$
while neutron EDM receive contributions after integrating out the
scalar leptoquark $\phi_{i1}$. This feature in this model allows
us to avoid the strong constraints imposed by neutron EDM on the
Yukawa couplings relevant to the transition $\tau^-\to s\bar
u\nu_\tau$ as we will show in the following. It should be remarked
that in the previous studies  both neutron EDM and $\tau^-\to
s\bar u\nu_\tau$ transition receive contributions from one
leptoquark. As a result, in those studies, the strong constraint
obtained from the neutron EDM suppresses considerably the
leptoquark contributions to the $\tau^-\to s\bar u\nu_\tau$
transition,

Integrating out  $\phi_{i1}$ yields tensor contributions to the
neutron EDM, which can be expressed by the effective Lagrangian
$\Lagr^{\rm EDM}_T $  \bea \label{eq:LT1} \Lagr^{\rm EDM}_T
&\equiv& C^{\rm EDM}_T (\bar{\tau} \sigma_{\mu \nu} P_R
\tau)(\bar{u} \sigma^{\mu \nu} P_R u)  + {\rm h.c.}, \eea

The renormalization group evolution~\cite{Jenkins:2013wua} of the
operator $(\bar{\tau} \sigma_{\mu \nu}R  \tau)(\bar{u} \sigma^{\mu
\nu} R u)$ can produce via insertion an up-quark EDM $d_u (\mu)$
\cite{Cirigliano:2017tqn}
 \be \Lagr_{\rm D}=- \frac{i}{2}  d_u (\mu) \bar{u} \sigma^{\mu
\nu} \gamma_5   u F_{\mu \nu}. \ee
Upon solving the RG following
~\cite{Bellucci:1981bs,Buchalla:1989we,Cirigliano:2017azj} we
obtain
 \bea d_u (\mu) &=&   \frac{e
\mtau}{\pi^2} \,  Im \, C^{\rm EDM}_T (\mu)   \log
\frac{\Lambda}{\mu}\nonumber\\ &\simeq& 3.6 \, {\rm Im} \, C^{\rm
EDM}_T (\mu) \log \frac{\Lambda}{\mu} \times 10^{-15} \, e \,
cm.\eea

From the experimental current  $90\%$ C.L. upper bound on neutron
EDM: $d_n= g_T^{u} (\mu) d_u (\mu) < 1.8 \times 10^{-26}\,e\,{\rm
cm}$~\cite{Baker:2006ts,Afach:2015sja,Abel:2020pzs}, one obtains a
strong constraint on ${\rm Im} C^{\rm EDM}_T (\mu)$. Thus, for a
value $\Lambda \gsim 100 {\rm GeV}$ and using   and the recent
lattice result~\cite{Bhattacharya:2015esa} $g_T^{u} (\mu =  2{\rm
GeV}) = - 0.204(11)(10)$,  one finds the  constraint \be |{\rm Im}
C^{\rm EDM}_T (\mu_\tau) | \lsim 8.8 \times 10^{-12}, \label{ctc}
\ee which yields \be \left| {\rm Im} (Y^{2\,*}_{ 31} Y^{4\,*}_{
13}) \right| \lsim 8.8\times
\big[\frac{m_{\phi_{i1}}}{GeV}\big]^2\times 10^{-12}. \label{EDMc}
\ee
It should be noted that, the above constraint is based on the
assumption that there are no other contributions to $d_n$ that can
cancel the effect of $C^{\rm EDM}_T$.  As can be seen from
Eq.(\ref{EDMc}) that $\left| {\rm Im} (Y^{2\,*}_{ 31} Y^{4\,*}_{
13})\right| \lsim 8.8 \times 10^{-6}$, for leptoquark masses
$m_{\phi_{i1}} \simeq {\cal O}(1)$ TeV. This result agrees with
the bound obtained in Eq.(31) in Ref.\cite{Cirigliano:2017tqn}.

Possible constraints on the parameter space of the model can be
obtained from the new contributions to the $D-\bar D$ mixing. This
can be done upon integrating out the scalar leptoquark $\phi_{i1}$
leading to the effective Lagrangian $\Lagr^{ D\bar D}_T $ \bea
\label{eq:LT1} \Lagr^{ D\bar D}_T &\equiv& C^{ D\bar D}_T
(\bar{\tau} \sigma_{\mu \nu} P_R \tau)(\bar{c} \sigma^{\mu \nu}
P_R u)  + {\rm h.c.}, \eea

The double insertion of the operator $(\bar{\tau} \sigma_{\mu \nu}
P_R \tau)(\bar{c} \sigma^{\mu \nu} P_R u)$ results in tensorial
contribution to the $D-\bar D$ mixing that is proportional to $C^{
D\bar D}_T =\frac{1}{4 m^2_{\phi_{i1}}} Y^{2\,*}_{ 31} Y^{4\,*}_{
23}$. Consequently, for a given leptoquark mass $m_{\phi_{i1}}$,
one can obtain a bound on the Yukaw product $Y^{2\,*}_{ 31}
Y^{4\,*}_{ 23}$.

The transition $\tau\to s \bar u v_\tau $ generating the decay
process $\tau\to K^0_S\pi^- (K^-\pi^0) \nu_\tau$ can be obtained
upon integrating the leptoquark $\phi_{i2}$. Doing so and after
using Fierz identities, we find that
\begin{eqnarray}
\label{eq:LT2} \Lagr_{eff}  =-\frac{4}{m^2_{\phi_{i2}}} Y^{2\,*}_{
A1} Y^{4\,*}_{ D 3} V^{\rm CKM*}_{ D 2} V_{ A3}^{\rm PMNS*}
\bigg[(\bar{\nu}_{\tau} P_R \tau)(\bar{s}  P_R u)
+\frac{1}{4}(\bar{\nu}_{\tau} \sigma_{\mu \nu} P_R \tau)(\bar{s}
\sigma^{\mu \nu} P_R u)    \ + \ {\rm h.c.}\bigg].
\end{eqnarray}
The Yukawa couplings $Y^{2}$ and $Y^{4}$ are generally complex. In
this case, the Wilson coefficients $C_i$, corresponding to the
operators $Q_i$ in Eq.(\ref{Qi})are given by
\begin{eqnarray}
 C_{V}&=&0,\,\,\,\,\,\,\,\,\,\,\,\,\,\,\,\,\,\,
 C_{S}=\frac{ 2\sqrt{2}\, Y^{2\,*}_{ A1} Y^{4\,*}_{
D 3} V^{\rm CKM*}_{ D 2} V_{ A3}^{\rm PMNS*}}{\, G_F V^*
_{us}\,m^2_{\phi_{i2}}},\,\,\,\,\,\,\,\,\,\,\,\,\,\,\,\,\,\,
C_{T}=\frac{1}{4} C_{S}.\label{leptq}
\end{eqnarray}
 Due to the strong constraint on $|Im(Y^{2\,*}_{ 31} Y^{4\,*}_{
13})|$ obtained from the EDM of the neutron, we can neglect the
contributions of $Y^{2\,*}_{ 31} Y^{4\,*}_{ 13}$ to ${\rm Im}
(C_{T})$. This is the case also for the Yukawa product $Y^{2\,*}_{
31} Y^{4\,*}_{ 23}$ which is expected to be suppressed by the
constraint from $D-\bar D$ mixing as can be seen from the
expression of $C^{ D\bar D}_T $ given before. On the other hand
and in order to maximize the CP asymmetry, we need large values of
$Im(C_{T})$. This can be achieved by choosing $A=2$ and $D=2$ in
Eq.(\ref{leptq}) which correspond to large values of $V_{ 23}^{\rm
PMNS*}$ and $V^{\rm CKM*}_{ 22}$ respectively. As a result, the
relevant parameter space to our processes includes the leptoquark
$\phi_{i2}$ mass and the Yukawa couplings product $Y^2_{21}
Y^4_{23}$.

%
\section{Results and Implications}

In this section, we discuss how our $SU(5)$ Leptoquark contributes
to the CP asymmetry of $\tau^- \to K^- \pi^0 \nu_\tau$ and $\tau^-
\to K_S^0 \pi^- \nu_\tau$. The CP asymmetries of these decay
processes are given in terms of the ${\rm Im}(C_T)$, as shown in
Eq.(\ref{asym}). From the discrepancy between the $1 \sigma$
measured value of $A_{\rm CP}(\tau^- \to K_S^0 \pi^- \nu_\tau)$
and the SM expectation, one finds that ${\rm Im}(C_T)$ is
constrained as follows: \be - 0.342 \leq {\rm Im}[C_T(m_\tau)]
\leq - 0.118. \label{Imb} \ee Here, the tensor Wilson coefficient
at scale $m_\tau$, $C_T(m_\tau)$, is defined in terms of tensor
Wilson coefficient at scale $m_{\phi_{i2}}$, $C_T(m_{\phi_{i2}})$,
through the renormalization group equation which  can be expressed
as \be C_T(m_\tau)=R_T(m_\tau,
m_{\phi_{i2}})\,C_{T}(m_{\phi_{i2}})\,, \ee where the evolution
function $R_{T}(m_\tau, m_{\phi_{i2}})$ at the leading logarithmic
approximation is given by \cite{Dorsner:2016wpm}  \be
R_{T}(m_\tau,m_{\phi_{i2}}) \equiv
\bigg[\frac{\alpha_s(m_b)}{\alpha_s(m_\tau)}\bigg]^{\frac{\gamma_{T}}
{2\beta_0^{(4)}}}\,\bigg[\frac{\alpha_s(m_t)}{\alpha_s(m_b)}\bigg]^{\frac{\gamma_{T}}{2\beta_0^{(5)}}}\,
\bigg[\frac{\alpha_s(m_{\phi_{i2}})}{\alpha_s(m_t)}\bigg]^{\frac{\gamma_{T}}{2\beta_0^{(6)}}}
\,, \ee here $\beta_0^{(n_f)}=11-2n_f/3$ represents the
leading-order coefficient of the QCD beta function, $n_f$ is the
number of active quark flavours, and
$\gamma_T=8/3$~\cite{Gracey:2000am} stands for the leading-order
anomalous dimensions of the tensor current.

With the parameter space selection, it is critical to check that
the predicted result of the branching ratio of $\tau^- \to K^-
\nu_\tau$ to ensure that it remains within the experimental
limits. The experimental results for the branching ratio of
$\tau^- \to K^- \nu_\tau$ is given as ${\cal B}_{\rm Exp.}=
(6.96\pm0.10)\times 10^{-3}$ \cite{Workman:2022ynf}. The SM
expectation for this branching ratio can be expressed as
\begin{equation}
{ {\cal B}_{SM}}\left[ {{\tau^-} \to K^- \nu_\tau } \right] =
\frac{{{m^3_{{\tau^-}}}}}{{16\pi }} G_F^2 {\tau
_{{\tau^-}}}f_{{K}}^2{\left| {V^*_{u s}} \right|^2} {\left( {1 -
\frac{{m_{K^-}^2}}{{m_{{\tau^-}}^2}}}
\right)^2}\left(1+\delta^{K}_{ RC}\right)\,,
\end{equation}
where $\delta^{K}_{ RC}$ accounts for the short and long distance
electromagnetic radiative corrections. Using the mean life $\tau
_{\tau^-} =(290.3\pm 0.5)\times 10^{-15} s$
\cite{Workman:2022ynf}, $\delta^{\tau K^-}_{ EM}=2.04(62)\%$
\cite{Cirigliano:2021yto} and the results of the latest global SM
analysis reported by the UTfit collaboration $|V_{us}|=0.2251(8)$
\cite{UTfit:2022hsi}, $f_{\pi}=92.3(1)$ MeV \cite{Workman:2022ynf}
and $f_K= 1.198 f_{\pi}$ \cite{Workman:2022ynf}.  Numerically it
leads to ${\cal B}_{SM} \sim 3.57 \times 10^{-3}$. The branching
ratio is modified by a New Physics (NP) contribution, such as
leptoquark contribution given in Eq.(\ref{eq:LT2}), as follows:
\begin{equation}
{{\cal B}_{NP}}= {\cal B}_{SM}  \bigg(1 +
\left|\frac{{m_{K}^2}}{\left(m_{u}+m_{s}\right) m_{\tau^-}} C
_S\right|^2\bigg) \label{eq:BrNP}
\end{equation}
In this regard, the Wilson coefficient $C_S = 4 C_T$ is
constrained by \be \vert C_S \vert \lsim \frac{(m_u + m_s)
m_\tau}{m_K^2} \left( \frac{{\cal B}_{Exp}}{{\cal B}_{SM}} - 1
\right)^{\frac{1}{2}} \simeq 0.67. \label{CS} \ee

\begin{figure*}[!ht]
\includegraphics[width=10cm,height=8cm,angle=0]{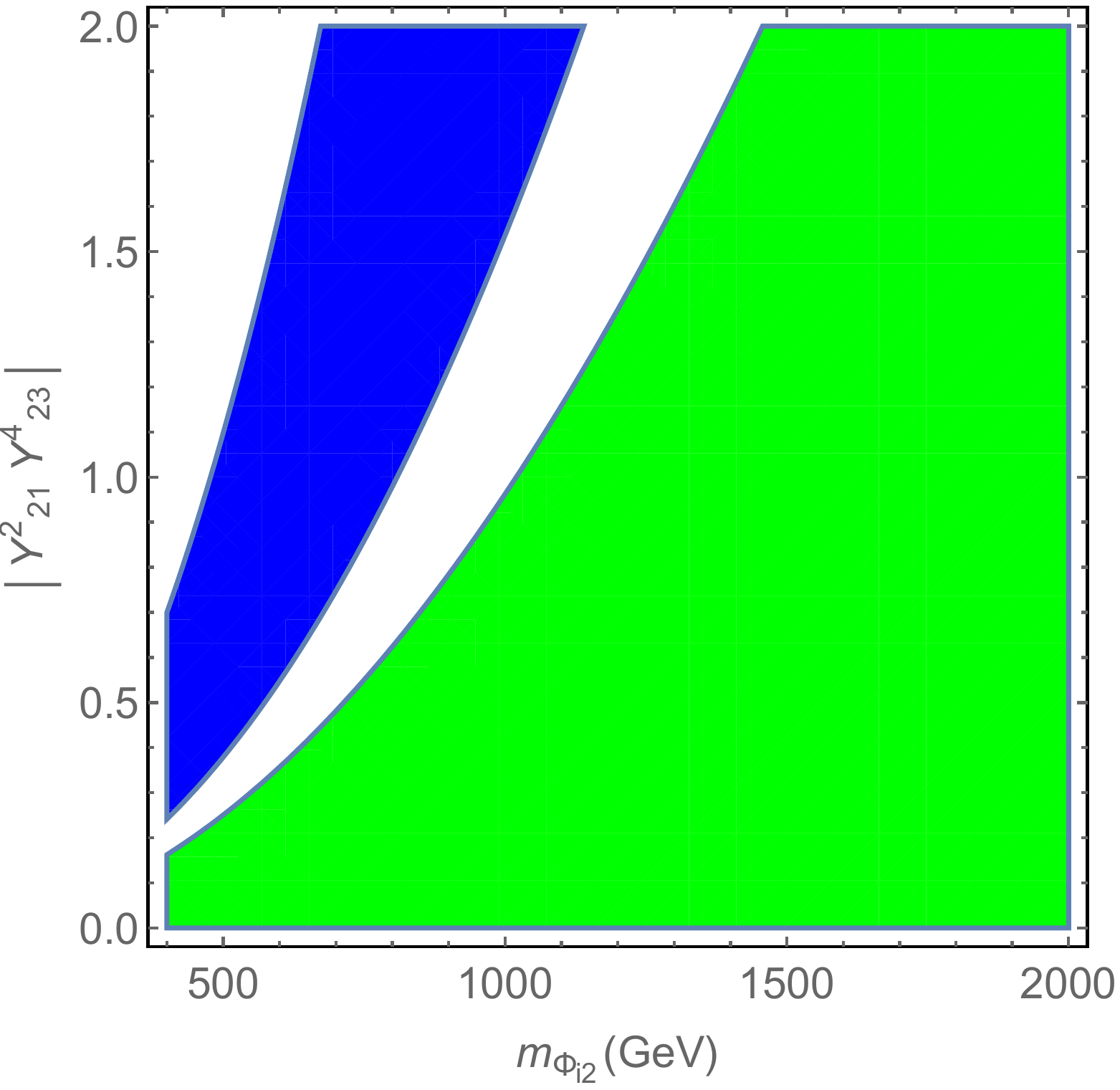}~~~~~~~~
\caption{ Allowed region of the parameter space $( m_{\phi_{i2}},
\vert Y^2_{21} Y^4_{23}\vert)$ in blue (Green) color satisfying
the obtained bound in Eq.(\ref{Imb}) (Eq.\ref{CS}). } \label{fig1}
\end{figure*}

In Fig. \ref{fig1}, we plot in blue (Green) the allowed region of
the parameter space $( m_{\phi_{i2}}, \vert Y^2_{21}
Y^4_{23}\vert)$ satisfying the  obtained $1\sigma$ bound in
Eq.(\ref{Imb}) (Eq.\ref{CS}). As shown in the figure, there is no
region in the $( m_{\phi_{i2}}, \vert Y^2_{21} Y^4_{23}\vert)$
parameter space that meets both constraints. Therefore, the
constraint from the branching ratio of of $\tau^- \to K^-
\nu_\tau$  is stringent at $1\sigma$ level, and the new
contributions of the scalar leptoquark cannot resolve the direct
CP asymmetry of the decay $\tau^- \to K_S^0 \pi^- \nu_\tau$.
However, if we relax the limits in Eq.(\ref{Imb}) to consider the
$2\sigma$ level, we find that these new contributions can
accommodate the mentioned CP asymmetry.

\begin{figure*}[!t]
\includegraphics[width=8.5cm,height=7cm,angle=0]{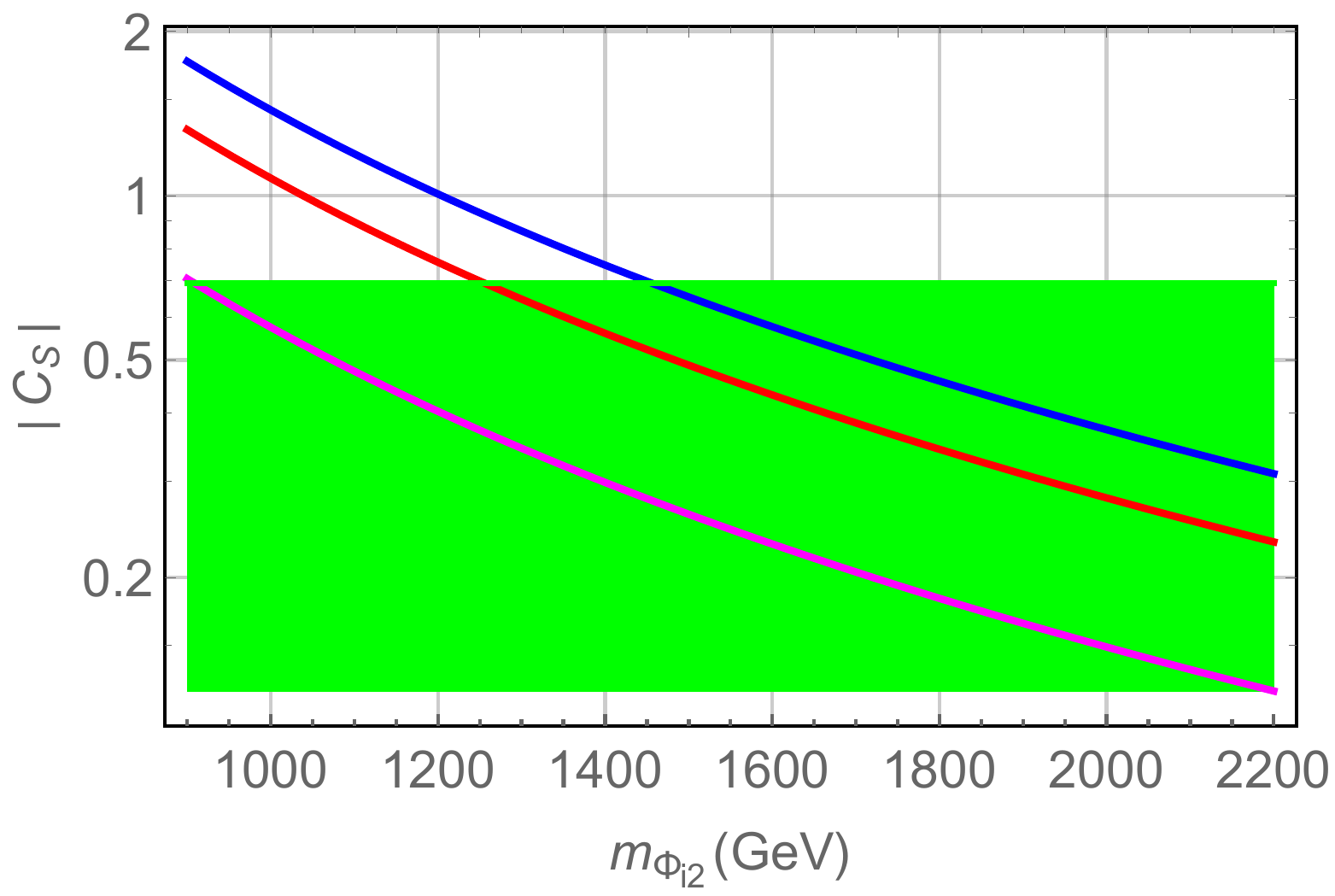}~~~~~~~~
\includegraphics[width=8.5cm,height=7cm,angle=0]{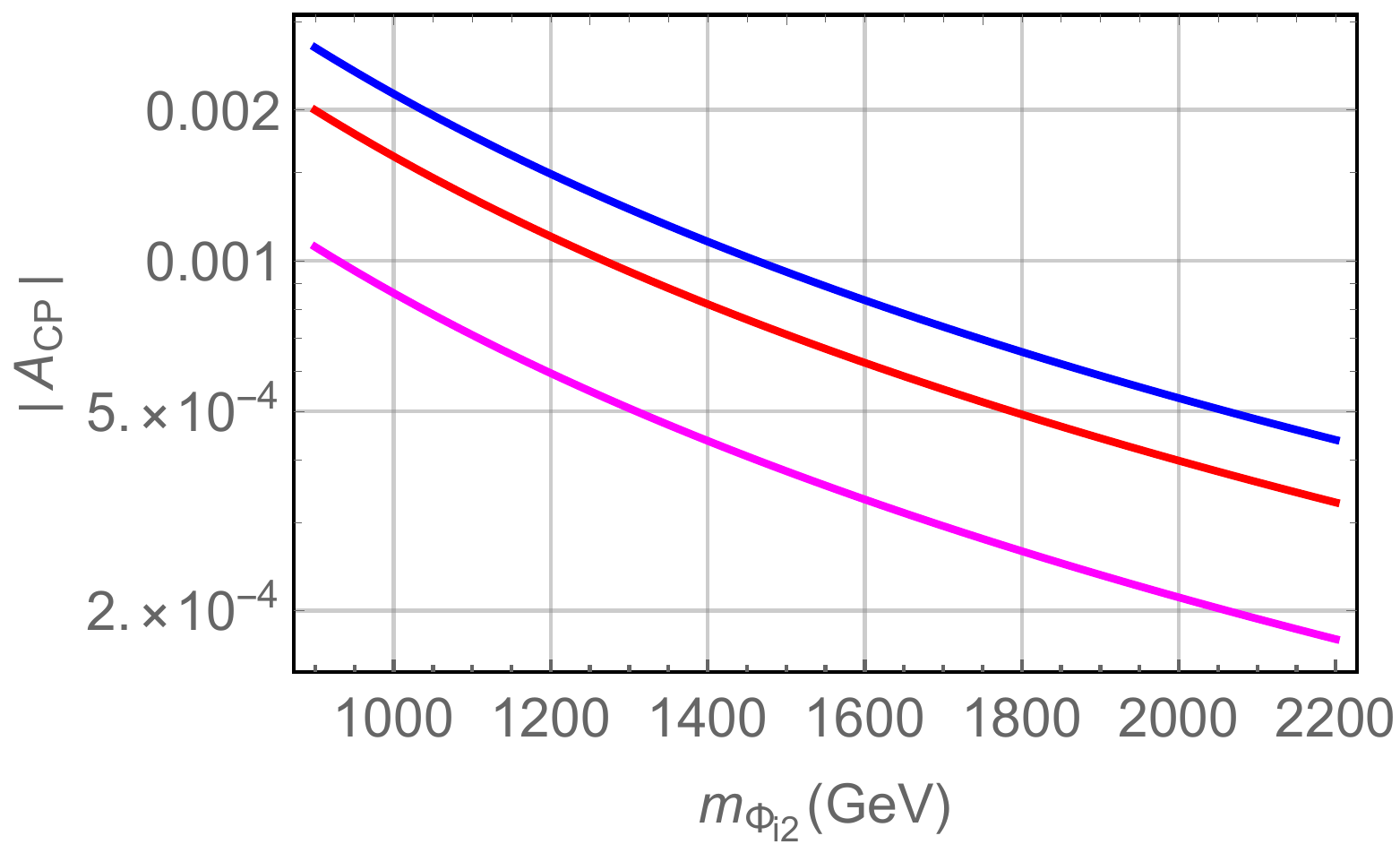}
\caption{$\vert C_S \vert$ ($\vert A_{CP} \vert$) of the process
$\tau^- \to K^-\pi^0 \nu_\tau$ left (right) as function of
leptoquark mass $\vert Y^2_{21} Y^4_{23}\vert =0.8,1.5,2.0$ in
magenta, red and blue colors respectively. The green region
represent the $1 \sigma$ bound in eq.(\ref{CS}). } \label{fig2}
\end{figure*}

\begin{figure*}[!ht]
    \includegraphics[width=10cm,height=7cm,angle=0]{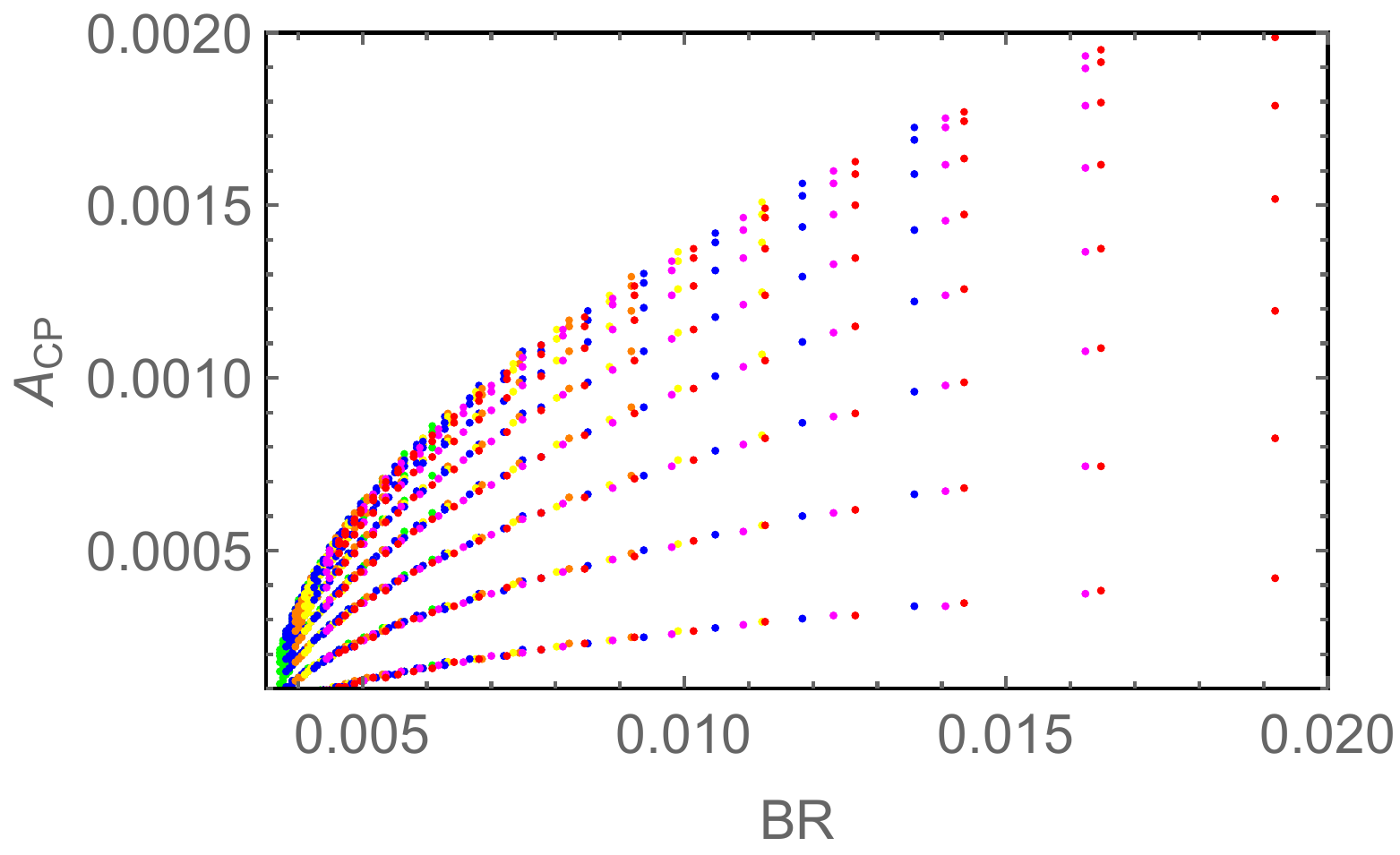}~~~~~~~~
    \caption{ Correlation of $BR={ {\cal B}_{NP}}\left[ {{\tau^-} \to
            K^- \nu_\tau } \right]$ and $\vert A_{\rm CP}(\tau \to K^- \pi \nu_\tau) \vert$ channel.
    }
    \label{correlation}
\end{figure*}

Our objective in this study is to give an estimation of the direct
CP asymmetry of the decay mode $\tau^- \to K^-\pi^0 \nu_\tau$. In
Fig. \ref{fig2}, we show in the left plot $\vert C_S \vert$ at
$m_\tau$ scale while in the right plot we display $\vert A_{CP}
\vert$ of the mode $\tau^- \to K^-\pi^0 \nu_\tau$ as a function of
leptoquark mass at fixed Yukaw couplings product $\vert Y^2_{21}
Y^4_{23}\vert =0.8,1.5,2.0$ in magenta, red and blue colors
respectively. It should be noted that, the green region represent
the $1 \sigma$ bound in eq.(\ref{CS}). The left plot can be used
to determine the value of the $\vert Y^2_{21} Y^4_{23}\vert$
corresponding to a given leptoquark mass, ensuring that the
branching ratio of the decay $\tau^- \to K^- \nu_\tau$ remains
consistent with its experimental value. From the right plot in
Fig. \ref{fig2}, for these allowed values of parameter space $(
m_{\phi_{i2}}, \vert Y^2_{21} Y^4_{23}\vert)$, we can deduce from
that $A_{\rm CP}(\tau \to K^- \pi \nu_\tau)$ could be in the
$10^{-3}$ to $10^{-4}$ range for a consistent branching ratio of
$\tau^- \to K^- \nu_\tau$. This conclusion is supported by
Fig.\ref{correlation}, which shows the relationship between the
$A_{\rm CP}(\tau \to K^- \pi \nu_\tau)$ and  ${\rm BR}(\tau \to
K^- \nu_\tau)$.

\section{Conclusion \label{sec:conclusion}}

In this paper, we have investigated the possibility of resolving
the $2.8~\sigma$ deviation between the SM prediction and the
experimental result of the CP asymmetry in the decay  $\tau^- \to
K_S^0 \pi^- \nu_\tau$,  using a low scale non-minimal $SU(5)$
mode. We emphasized that the non-minimal SU(5), in which the Higgs
sector is extended by an extra 45-dimensional multiplet, provides
a TeV scalar triplet leptoquark that generates direct CP asymmetry
in semi-leptonic $\tau$ decay $\tau^- \to K_S^0 \pi^- \nu_\tau$,
and thus can account for this discrepancy. Furthermore, we
demonstrated that, within the same parameter space, the CP
asymmetry of $\tau^- \to K^- \pi^0 \nu_\tau$ is of order ${\cal O}
(10^{-3})$, which is a few orders of magnitude larger than the
results obtained in other models beyond the SM.  This can be
attributed to the presence of Yukawa couplings of our leptoquark
that do not contribute to proton decay,  EDM  and $D-\bar D$
mixing, enabling us to avoid the related constraints. The phases
of these couplings allow us to enhance the CP asymmetries
considerably. We also demonstrated that the branching ratio of
$\tau \to K^- \nu_\tau$, after including the leptoquark
contributions with large Yukawa couplings required to enhance the
CP asymmetry $A_{\rm CP}(\tau \to K_S^0 \pi^- \nu_\tau)$, is well
within the experimental limits.


\section*{Acknowledgements}
The work of S. K. is partially supported by Science, Technology $\&$ Innovation Funding Authority (STDF) under grant number 37272.


\end{document}